# New Radio Observations of the Supernova Remnant CTA 1

Tam Do[1,2], Roland Kothes[2], Alex S. Hill[2,3], Andrew Gray[2], Patricia Reich[4], and Wolfgang Reich[4]
[1] Department of Physics and Astronomy, University of British Columbia, 6224 Agricultural Road, Vancouver, BC V6T 1Z1, Canada
[2] Dominion Radio Astrophysical Observatory, Herzberg Astronomy and Astrophysics Research Centre, National Research Council Canada Box 248, Penticton, BC V2A 6K3, Canada
[3] Department of Computer Science, Math, Physics, and Statistics, University of British Columbia–Okanagan, 1177 Research Rd, Kelowna, BC V1V 1V7, Canada
[4] Max-Planck-Institut für Radioastronomie, Auf dem Hügel 69, 53121 Bonn, Germany
Received 2023 September 14; revised 2024 October 27; accepted 2024 November 7; published 2024 December 19

## Abstract

We present new radio images of the supernova remnant (SNR) CTA 1 at 1420 and 408 MHz, and in the 21 cm line of H I observed with the Dominion Radio Astrophysical Observatory Synthesis Telescope and at 1420 MHz observed with the Effelsberg 100 m telescope. We confirm previously described continuum features and elaborate further on filamentary features identified using the high-resolution (1′) maps from these new observations. We investigate the abrupt change in sign of rotation measure (RM) across the SNR, using the linear polarization observations in the four bands around 1420 MHz. Following X. H. Sun et al.'s (2011) investigation, we both confirm that the distribution of signs of the RMs for extragalactic sources in the area appears to match that of the shell, as well as combine the data from the four bands to estimate the relative depolarization and the intrinsic rotation measure of the SNR. We do not conclusively reject X. H. Sun et al.'s (2011) claim of a Faraday screen in the foreground causing the distribution of RMs that we observe; however, we do suggest an alternative explanation of a swept-up stellar wind from the progenitor star with a toroidal magnetic field. Finally, we expand on the analysis of the H I observations by applying the Rolling Hough Transform to isolate filamentary structure and better identify H I emission with the SNR. Further constraining the H I velocity channels associated with CTA 1, we use more recent Galactic rotation curves to calculate an updated kinematic distance of $1.09 \pm 0.2$ kpc.

*Unified Astronomy Thesaurus concepts:* Supernova remnants (1667)

## 1. Introduction

The supernova remnant (SNR) CTA 1 (G 119.5 + 10.3) was first discovered by D. E. Harris & J. A. Roberts (1960) and has since been observed over a wide range of wavelengths. At radio frequencies, high-angular resolution continuum maps have been made by the Effelsberg 100 m telescope at 1720 and 2695 MHz (W. Sieber et al. 1979, 1981) and the DRAO Synthesis telescope combined with Effelsberg 100 m observations at 408 and 1420 MHz (S. Pineault et al. 1993, 1997). The general structure and morphology of the SNR could be discerned from the continuum emission and all observations indicate that CTA 1 is an incomplete shell SNR, breaking out into a region of low density, in the adiabatic phase of expansion (A. Bamba & B. J. Williams 2022). H I observations of angular resolution 5′, covering an effective range of velocities from +50 to −100 km s$^{-1}$, observed with the 4-antenna Synthesis Telescope and short-spacings additions from the single-dish 26 m telescope at DRAO are also presented in S. Pineault et al. (1993). They identify H I features potentially associated with CTA 1, most notably a completion of the continuum shell in the breakout region, which allowed for an estimate of the kinematic distance to the SNR. X. H. Sun et al. (2011) presented 4800 and 2639 MHz observations that allowed for rotation measure (RM) mapping. They suggest that an abrupt transition between positive and negative RMs across the SNR's shell, corroborated by RM patterns of extragalactic sources in the area, is indicative of a large Faraday screen (P. C. Tribble 1991) with a regular magnetic field in the opposite direction to the ISM interstellar medium field, partly covering CTA 1.

Its signature features in nonradio frequencies are strong [O III] filaments in the optical band, which coincide with the radio shell (R. A. Fesen et al. 1981, 1983; F. Mavromatakis et al. 2000), a radio-quiet gamma-ray central pulsar ($\alpha = 00^{\mathrm{h}}07^{\mathrm{m}}01\overset{s}{.}56$; $\delta = +73°03'08\overset{''}{.}10$) discovered with the Fermi-LAT (A. A. Abdo et al. 2008), a pulsar wind nebula detected with VERITAS (E. Aliu et al. 2013), strong center-filled X-ray emission consisting of a thermal component heated by reverse shock and a no-thermal component powered by the pulsar (F. D. Seward et al. 1995; P. Slane et al. 1997), and pulsations in the X-ray from the central pulsar detected with long XMM-Newton observations (P. A. Caraveo et al. 2010).

In Section 2, we describe the new observations and the data processing/analysis applied to them using established methods. In Section 3, we first present our new low-noise continuum maps at 1420 and 408 MHz, confirming features previously discussed in S. Pineault et al. (1993, 1997) and clarifying the filamentary structure within the shell that is identifiable with our clearer maps. We then present new linear polarization maps at 1420 MHz, and finally, we present new H I observations of higher spectral resolution from the 7-antenna Synthesis Telescope and short-spacings additions from the Effelsberg 100 m single-dish telescope, as well as our application of the Rolling Hough Transform (S. E. Clark et al. 2014) to the spectral line data. We confirm the association of the H I partial shell with CTA 1 and calculate an updated kinematic distance using the newer Galactic rotation curve from M. J. Reid et al. (2019) and reexamine the possible associations of other features. We discuss the possible implications of our analysis in

Table 1
Observation Characteristics of the DRAO ST Fields

| Field | R.A. (J2000) (HH MM SS.SS) | Decl. (J2000) (deg arcmin arcsec) | Galactic Longitude (deg) | Galactic Latitude (deg) | First obs. (YYYY-MM-DD) | Last obs. (YYYY-MM-DD) | Noise H I (K) | Noise C21 (K) | Noise C74 (K) |
|---|---|---|---|---|---|---|---|---|---|
| DA0  | 00 13 16.55 | +70 05 58.4 | 119.67 | +7.47  | 2017-10-2  | 2017-12-5  | 3.0 | 0.045 | 1.00 |
| DA1  | 23 41 11.70 | +72 51 30.0 | 117.74 | +10.70 | 2017-05-8  | 2017-10-12 | 3.0 | 0.046 | 0.91 |
| DA3  | 23 57 55.70 | +71 30 30.0 | 118.67 | +9.08  | 2017-07-22 | 2017-10-21 | 2.9 | 0.041 | 0.90 |
| DA5  | 23 51 16.53 | +69 43 54.1 | 117.73 | +7.47  | 2004-04-24 | 2004-06-9  | 2.7 | 0.043 | 0.96 |
| DA6  | 00 10 13.60 | +71 41 30.0 | 119.67 | +9.08  | 2017-05-9  | 2017-10-1  | 3.0 | 0.046 | 1.00 |
| DA8  | 23 35 36.64 | +71 03 12.0 | 116.80 | +9.09  | 2004-04-21 | 2004-06-8  | 3.0 | 0.046 | 1.11 |
| DA9  | 00 06 35.66 | +73 17 13.1 | 119.67 | +10.70 | 2017-10-1  | 2017-12-12 | 3.0 | 0.044 | 0.91 |
| FR07 | 00 34 56.49 | +70 17 12.1 | 121.53 | +7.46  | 2019-01-1  | 2019-04-2  | 2.1 | 0.041 | 1.05 |
| FR13 | 00 21 06.45 | +71 48 02.2 | 120.54 | +9.07  | 2019-05-11 | 2019-07-9  | 2.3 | 0.041 | 0.88 |
| FR14 | 00 44 47.33 | +71 56 05.6 | 122.41 | +9.07  | 2019-05-14 | 2019-07-8  | 2.2 | 0.041 | 0.90 |
| FR20 | 00 31 52.02 | +73 30 28.6 | 121.52 | +10.69 | 2019-11-27 | 2020-08-29 | 2.0 | 0.044 | 0.98 |
| FR21 | 00 57 39.62 | +73 33 22.3 | 123.38 | +10.69 | 2019-05-12 | 2019-07-13 | 2.4 | 0.041 | 1.00 |

Section 4, including our attempt to investigate the RM sign distribution described by X. H. Sun et al. (2011), and we summarize our findings and conclusions in Section 5.

## 2. Observations and Data Analysis

Observations for this project were obtained with the Synthesis Telescope (ST) at the Dominion Radio Astrophysical Observatory (DRAO). The ST is an east–west interferometer comprising seven 9 m diameter antennas with a minimum baseline of 12.9 m and a maximum baseline of 617.2 m. Relevant to the polarimetry, two of the antennas have a diameter of 9.14 m and a focal length of 3.81 m, while the remainder have a diameter of 8.53 m and a focal length of 3.66 m. With uniform weighting the synthesized beam has a half-power beamwidth of 0.′82 at 1420 MHz and 2.′8 at 408 MHz (elongated by cosec $\delta$ in the N-S direction at decl. $\delta$; T. L. Landecker et al. 2000). An elliptical Gaussian taper was applied to the u-v plane for the H I data, falling to 20% at the longest baseline, which increases sensitivity at a slight loss of resolution (to 58″).

The telescope simultaneously records data in both hands of circular polarization in the 21 cm H I line and adjacent radio continuum bands, and in right-hand circular polarization only in a continuum band at 74 cm (408 MHz). Data for each pointing center are acquired in 12 array configurations, which populate the u-v plane with concentric tracks, ranging from the minimum to maximum baselines at an increment of 4.3 m, or roughly half the antenna diameter. The resulting ensemble of data is referred to as a "field," with each field having a unique alphanumeric label known as the "field code." The well-filled u-v plane in a completed field yields dirty images after Fourier inversion that are characterized by very low sidelobes, with a single prominent grating lobe that falls close to the first null of the primary antenna response for a source at the field center.

The 21 cm system operates over a 35 MHz band centered on the H I line under observation: the central 5 MHz is filtered out and fed to a 256-channel correlation spectrometer, where both copolar spectra are formed, while the remaining two 15 MHz bands are filtered into pairs of 7.5 MHz bands, with all four co- and cross-polar quantities correlated in each band, allowing full recovery of Stokes I, Q, U, and V. The 74 cm system has a single 3.5 MHz band, producing a single copolar correlation product. Further specifications and capabilities of the ST can be found in T. L. Landecker et al. (2000) and R. Kothes et al. (2010).

To increase sensitivity to extended structures, multiple overlapping fields were used in this project, divided into two sets, the first denoted by the field code prefix "DA" and the second by the field code prefix "FR." Table 1 lists the parameters for the individual fields. Note that the DA and FR fields used different spectrometer settings:

1. DA fields used a 1 MHz bandwidth centered at $v_{\rm LSR} = -60\,{\rm km\,s^{-1}}$, with a resulting channel width of $0.83\,{\rm km\,s^{-1}}$ and a velocity resolution of $1.3\,{\rm km\,s^{-1}}$;
2. FR fields used a 2 MHz bandwidth centered at $v_{\rm LSR} = -100\,{\rm km\,s^{-1}}$, with a resulting channel width of $1.65\,{\rm km\,s^{-1}}$ and a velocity resolution of $2.6\,{\rm km\,s^{-1}}$.

The 1$\sigma$ rms noise in the emission-free spectrometer channels is given for each field in Table 1.

All fields were subjected to standard data processing steps (A. G. Willis 1999) and calibration procedures that were developed for the Canadian Galactic Plane Survey and are described in A. R. Taylor et al. (2003) for continuum at 1420 MHz and the H I data, T. L. Landecker et al. (2010) for the linearly polarized component at 1420 MHz, and A. K. Tung et al. (2017) for the 408 MHz continuum data.

Each field was masked beyond the 20% response level of the primary beam of the individual antennas (a radius of 93′ at 1420 MHz and 280′ at 408 MHz), then corrected for the primary-beam pattern prior to mosaicking. The twelve fields were mosaicked together with each one weighted proportionally to the inverse of its relative rms noise, and points within each field assigned a weight proportional to the square of the primary beam pattern at each position.

The final step was to recover missing spatial structures (so-called "short spacings") to which the ST is not sensitive (larger than about 40′ at 1420 MHz and 2° at 408 MHz). For this, we used single-antenna observations from the Effelsberg 100 m telescope, which give excellent spatial-frequency overlap with the ST, providing a smooth transition between structures imaged by the interferometer and the 100 m Effelsberg dish. ST and Effelsberg images were merged after tapering in the overlap region of the respective Fourier domains (A. R. Taylor et al. 2003; T. L. Landecker et al. 2010). The resulting short-spacings-added mosaic contains no negative "bowls" around any extended emission down to the level of the noise, giving

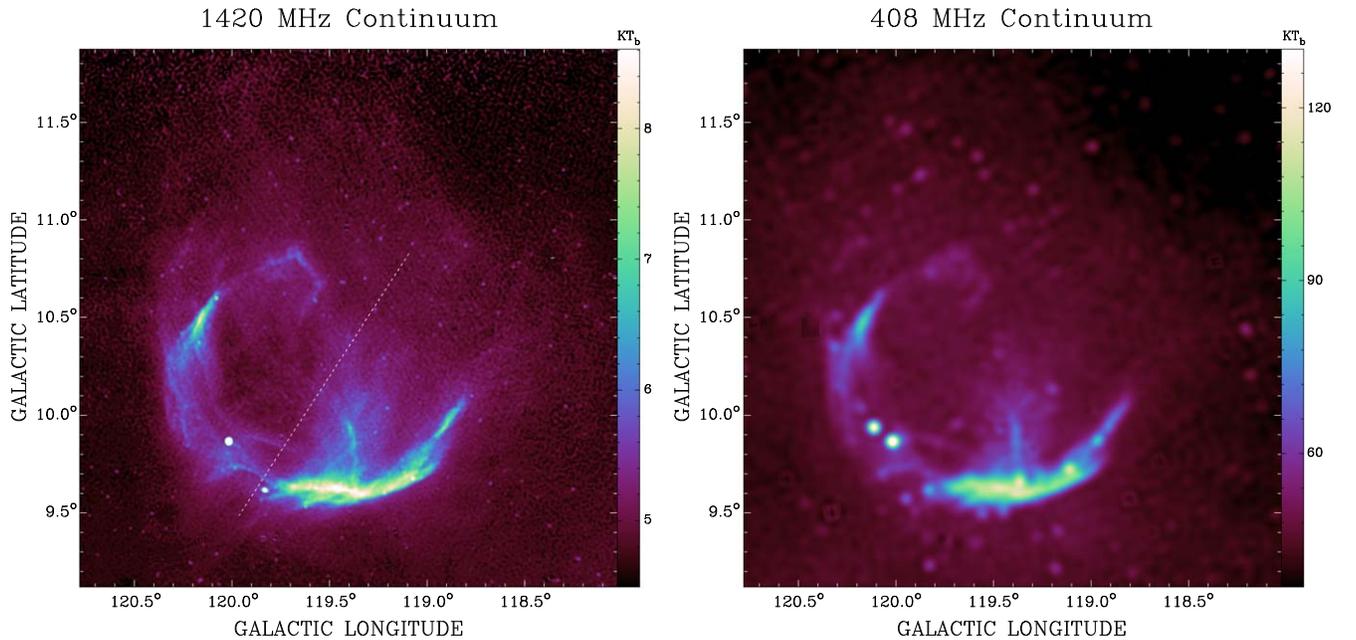

**Figure 1.** Continuum map of CTA 1 at 1420 (left) and 408 MHz (right) obtained with the Synthesis Telescope, after correcting for primary beam attenuation of the individual fields, and short-spacing data added from the EMLS and the Haslam 408 MHz Survey. The dashed white line indicates the symmetry axis for the possible stellar wind interpretation in Section 4.2. The angle of the bilateral symmetry axis was taken from J. L. West et al. (2016). The resolution is about 49″ at 1420 MHz and 2′.8 at 408 MHz. The color gradient shows the brightness temperature in Kelvin.

confidence not only in our short-spacing addition but in our flux calibration as well.

The short-spacing data set we used for our H I observations was the Effelsberg-Bonn H I Survey (EBHIS; B. Winkel et al. 2016), a survey with a spatial resolution of 10.′82 and a noise per 1.29 km s$^{-1}$ channel of $\Delta T = 0.09$ K. For radio continuum and polarization images at 1420 MHz we used data from an unpublished section of the Effelsberg 1.4 GHz Medium Galactic Latitude Survey (EMLS; W. Reich et al. 2004), and at 408 MHz we used the Haslam 408 MHz Survey (C. G. T. Haslam et al. 1982). The EMLS data are put on an absolute zero level by using the Stockert 25 m 1420 MHz Survey of the Northern sky (W. Reich 1982) for total-power intensities and the Low-Resolution DRAO 26 m Survey of polarized emission at 1.4 GHz for the linear polarization data (M. Wolleben et al. 2006).

### 3. Results

#### 3.1. CTA 1 in Radio Continuum

The full-resolution total-power maps at 1420 and 408 MHz are shown in Figure 1. We used the program `fluxfit` from the DRAO export software package (L. A. Higgs et al. 1997) to remove some bright point sources via Gaussian fitting. However, some of those compact sources were difficult to remove without affecting the shell emission from CTA 1, such as the planetary nebula NGC 40 at 120°.0 Galactic longitude and 9°.9 Galactic latitude. In some cases, the sources may be resolved and therefore difficult to fit with a Gaussian distribution, whereas in other cases, the background around those sources may have been too complex. The signal-to-noise ratio across the maps is high overall, as 12 observed fields were combined with noise-specific weights, with slightly higher noise toward the north as we have less overlapping field coverage there (see Table 1). The short-spacing additions from the Effelsberg observations (1420 MHz) and Jodrell Bank observations (408 MHz) provide information on the diffuse extended structure beyond the compact shell.

The continuum-emission features identified by S. Pineault et al. (1993) and S. Pineault et al. (1997) are seen clearly in the 1420 MHz map. Starting in the east, we see the bright eastern component of the shell splitting in the north, with some emission continuing nearly straight north along the edge of the broader emission seen best in the EMLS data and some emission along a curve inwards to connect with the "hook" feature at 119°.7 Galactic longitude and 10°.75 Galactic latitude. In the southeast portion of the shell, we confirm the "reverse shell" feature with a center of curvature at around 120° Galactic longitude and 9°.55 Galactic latitude. There appears to be a small section of the main shell that is of opposite curvature, which S. Pineault et al. (1997) interpret as an overdense region causing a decrease in shock velocity at that point. The "central emission bridge," which is thick and nearly vertical, extending from the southern shell toward the center of the system, can be seen to have stronger filaments highlighted in the broader column, as seen in the contour map from S. Pineault et al. (1997). However, the convolved 5-resolution map from S. Pineault et al. (1993) shows a "tongue" of emission extending beyond the circular boundary of the bright shell. This feature is not seen in the continuum maps from S. Pineault et al. (1997) and consistently, it is not seen in our maps either.

Finally, with a more sensitive map at 1420 MHz, the filamentary structure of the continuum shell is more distinguishable. We see this most in the eastern component, where instead of the shell being one uniformly bright broad filament, as is in the southwest component, we see two separate filaments making up the shell structure; one broad outer filament and a thinner inner filament, projected to the plane of the sky. The inner filament makes less of a rounded shape than the outer filament, with one straighter segment connecting the central emission bridge and eastern shell component from (+119°.6, $+9°.75$) to ($+120°$, $+9°.9$) and another segment that is separated from the outer filament by a small gap but traces the same shape from ($+120°$, $+9°.9$) to ($+120°.2$, $+10°.35$).

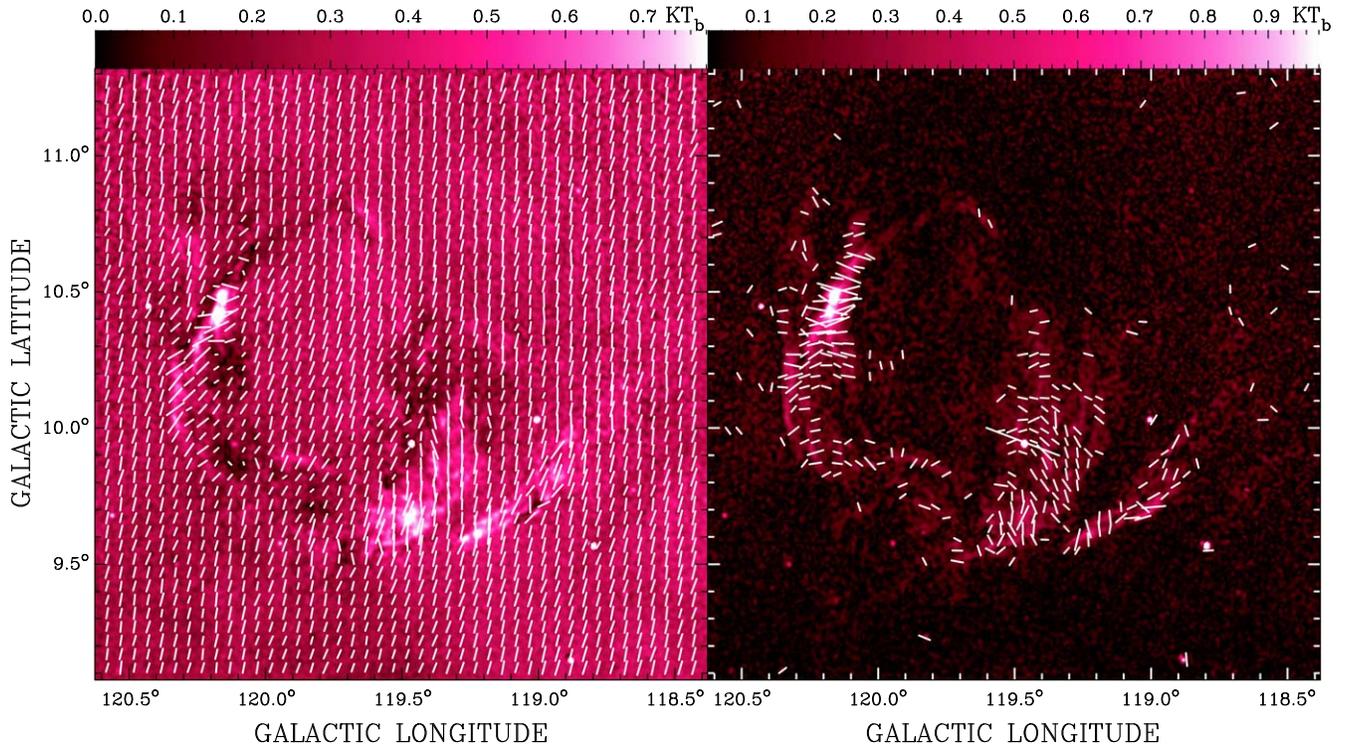

**Figure 2.** Linearly polarized emission from CTA 1 at 1420 MHz, obtained using the DRAO Synthesis Telescope, with short-spacing data added from the EMLS (left) and without short-spacing data (right). Vectors parallel to the electric field indicate the angle of the polarized emission, with the magnitude of the intensity shown by both the color gradient of the image in Kelvins of brightness temperature and length of the vector. In the right image, vectors are shown only for a polarized intensity above 0.15 K, which corresponds to about $5\sigma$. The resolution is $1'$ and the color gradient shows brightness temperature in Kelvin.

### 3.2. Linear Polarization and Faraday Rotation

We present polarized intensity maps at 1420 MHz, overlaid with vectors showing the polarization angle (aligned with the electric field vector) and magnitude of the polarization intensity in Figure 2. The image in the left panel contains the short spacing information from the EMLS, while in the right panel, we show the DRAO ST observations only. CTA 1 sits on top of the so-called Fan Region, an area of bright and smooth polarized emission where the polarization angle changes very slowly with frequency (R. G. Bingham & J. R. Shakeshaft 1967; W. N. Brouw & T. A. T. Spoelstra 1976). In this direction, the polarized emission from the Fan Region is clearly dominating the overall synchrotron emission. In the DRAO ST–only image, in the right panel of Figure 2, only features smaller than about $40'$ are visible since the DRAO ST is not sensitive to larger structures at this frequency (see also Section 2), naturally filtering out the smooth emission from the Fan Region. For our study, we are using the DRAO ST images, as we are interested in the thin filaments of the SNR. The removal of smooth emission coming from the SNR does not affect our polarization analysis. We see in Figure 2, right panel, that the magnetic field (oriented perpendicular to the plotted electric field) is generally aligned with the shell of the SNR, which is expected for little to no Faraday rotation, although the southwest part of the shell is slightly more disorganized. We also see a clear break across the shell from ($+119°.5$, $+9°.5$) to ($+119°$, $+10°.2$) where the shell appears depolarized, although this feature is notably not present in the continuum emission at both 1420 and 408 MHz.

### 3.3. Rolling Hough Transform Analysis of the H I Environment

The Synthesis Telescope and Effelsberg 100 m telescope H I observations cover the velocity range of $+45.5$ to $-164.7$ km s$^{-1}$. We use a mosaic of 12 observed fields (Table 1) to create a low-noise, high-resolution data cube, which we use to reexamine the features of interest from the last published analysis of the H I environment around CTA 1 by S. Pineault et al. (1993). We display the data in the upper panel of Figure 3.

To follow up on possible associations between the H I features and CTA 1 and more tightly constrain the feature's velocity ranges, we employ the Rolling Hough Transform (RHT; S. E. Clark et al. 2014). There are other filament/edge-detection methods, such as the algorithm developed for solar images by M. Qu et al. (2005) or basic unsharp masking; however, we choose the RHT as it was developed specifically to quantify the linearity and spatial coherence of H I structures and their alignment with the interstellar magnetic field. We use this method to isolate H I filaments in each velocity channel to create a corresponding H I filament cube. We show the backprojection of the RHT function (the integration of RHT intensity over all angles, normalized to be a probability from 0 to 1 for each pixel being part of a linear structure) for three velocity ranges of interest in Figure 3.

The first feature noted by S. Pineault et al. (1993) that we investigate is at low velocity: an H I bar running northwest to southeast, coincident with the hook feature in the north. We constrain this feature to three velocity channels spanning the range $-2.30$ to $-3.95$ km s$^{-1}$. Although we see what appears to be a bar in the H I channels, as described previously, the RHT-determined filamentary structure shows us that the gas is not likely connected and the filament coincident with the hook is not similarly shaped so the feature is not likely to be related.

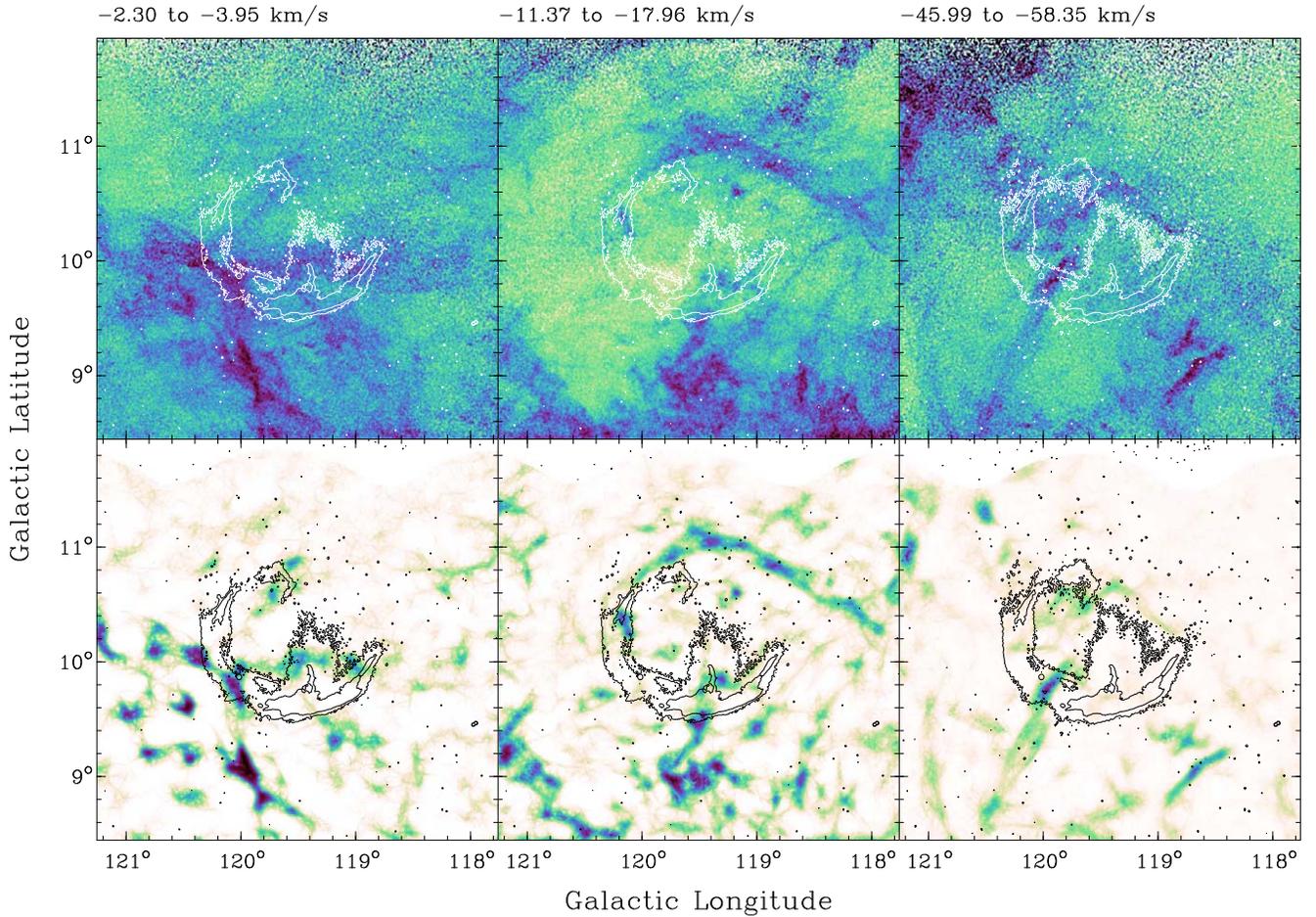

**Figure 3.** H I emission around CTA 1 (top row) obtained with the Synthesis Telescope with short-spacing data added from the EBHIS, and the RHT backprojection (bottom row) showing the probability in color intensity that H I emission is part of a coherent filamentary structure. The H I emission intensities, from left to right, range from 6.3–68.44, 2.94–43.88, and 2.75–14.84 in brightness temperature (K), with higher temperature in dark blue and lower temperature in green/yellow. The range of probabilities shown in the RHT backprojections, from left to right, range from 0–0.974, 0–0.847, and 0–0.749 with the highest probabilities in dark blue and the lowest probabilities in white. Contours of the 1420 MHz continuum data are shown at 5.3 and 6.1 K for reference in white on the top row and black on the bottom row, to indicate the location of CTA 1.

In the significantly higher velocity range of −46 to −58 km s$^{-1}$, S. Pineault et al. (1993) noted that there is H I emission wrapping around the north end of the central emission bridge. We observe this feature clearly in both the newer H I data and RHT filament structures, and we see this filament continues along the inside of the hook feature as well. However, we could not think of a convincing explanation for that morphology of H I gas being associated with an SNR, so we err on the conservative side and assume it is an unrelated coincidence.

The most convincing related feature, an apparent completion of the continuum shell in the northwest, lies in the mid-range velocity channels. S. Pineault et al. (1993) constrain the velocity range for the H I shell to −12 to −20 km s$^{-1}$. Using our higher resolution data and consulting the RHT filament image, we further constrain the range to −11.37 to −17.96 km s$^{-1}$, which corresponds to nine velocity channels (Figure 4). A small, bright H I cloud, coincident with an IRAS far-infrared point source at (+119°.2, +10°.6) is identified by S. Pineault et al. (1993) in the same velocity range, and we confirm that it persists in brightness throughout that range in the newer H I data as well. As this places the cloud at the same kinematic distance as the shell (and what we assume is the distance of the SNR), it suggests a possible relationship between the SNR and the IRAS point source.

## 4. Discussion

### 4.1. RHT Method for SNR Analysis

As mentioned in Section 3.3, we applied the RHT method to the H I emission surrounding CTA 1. The RHT was designed to roll over a single image, encoding the probability that any given pixel is part of a coherent linear structure, quantifying regional linearity without the specification of individual discrete fibers. We first applied it to all velocity-channel images from our data cube that contained a significant amount of H I emission, using the same window length and smoothing radius for all channels. We then compared the RHT backprojection images with the original H I emission images by eye and adjusted the smoothing radius and window-length parameters as necessary to ensure all possible structures were identified by the RHT function.

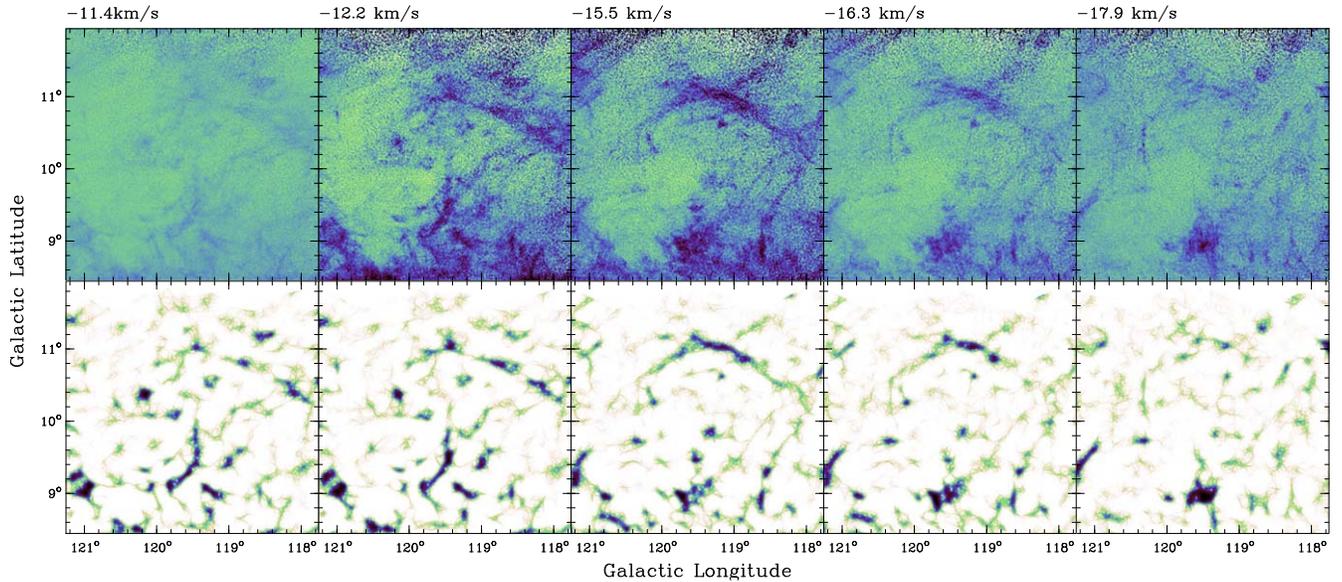

**Figure 4.** Comparisons between H I emission (top row) and RHT identified filamentary structure (bottom row) for five of the nine velocity channels in the range where the completion of the SNR's shell in H I and bright cloud coincident with the IRAS point source at (119°.2, 10°.6) are present.

Using the RHT method on H I emission around SNRs could make it easier to identify associated H I structures and constrain their range of velocities, thereby constraining the estimate of their kinematic distances. The RHT also records angular information about the linear structures and in future investigation, comparing the angle of linear structures with the linear polarization angles could lend insight into the magnetic field around CTA 1.

With an improved estimate of the SNR's velocity range, we recalculate the kinematic distance to the shell seen in the continuum as well as the identified IRAS far-infrared point source. For our calculation, we use the same standard International Astronomical Union (IAU) flat Galactic rotation curves used by S. Pineault et al. (1993), at the longitude of +119°.5, as well as the most recent rotation curves calculated by M. J. Reid et al. (2019). The updated kinematic distance fits are shown in Figure 5, alongside the kinematic distances calculated previously by S. Pineault et al. (1993). Their reported value was $1.4 \pm 0.3$ kpc, using a systemic velocity of $-16$ km s$^{-1}$ (the center of their constrained velocity range) and an uncertainty of two single-channel widths. We find, using the same IAU standard curve, a distance of $1.22 \pm 0.2$ kpc, and for the curve from M. J. Reid et al. (2019), we calculate a distance of $1.09 \pm 0.2$ kpc. We choose a systemic velocity of $-13.8$ km s$^{-1}$ since that is where the shell is the brightest and most structured (as observed from the new H I data and RHT filament maps). For uncertainty, we approach the determination of the velocity range for the shell in two ways. The first range we consider is the range of velocity channels over which the shell appears, as shown in Figure 4. We then make a first moment map of the H I emission over that range to get the intensity-weighted velocity, as shown in Figure 6. We determined that the range of velocities within the outline of the shell completion is 2.2 km s$^{-1}$ and then combined this range

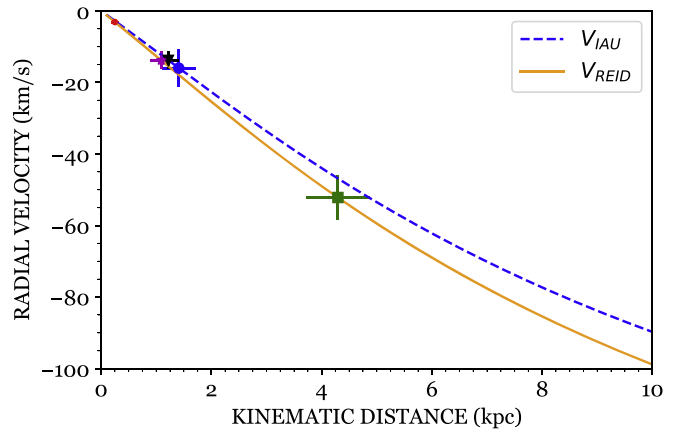

**Figure 5.** Kinematic distance calculated using H I velocities and Galactic rotation curves from M. J. Reid et al. (2019; orange solid line) and the IAU standard curve (blue dashed line). The red and green markers show, respectively, the distance for the low-velocity range ($-2.30$ to $-3.95$ km s$^{-1}$) and high-velocity range ($-46$ to $-58$ km s$^{-1}$) discussed in Section 3. The blue marker shows the distance for the range of the SNR shell calculated by S. Pineault et al. (1993) using the IAU rotation curve and systemic velocity of $-16$ km s$^{-1}$. The black triangle and purple star show, respectively, the distance for our new systemic velocity of $-13.8$ km s$^{-1}$ using the IAU rotation curve and M. J. Reid et al.'s (2019) curve, respectively.

with that of the span of channels, 6.59 km s$^{-1}$, using an rms:

$$\text{Combined Velocity Range: } \sqrt{\frac{6.59^2 + 2.2^2}{2}} \text{ km s}^{-1}. \quad (1)$$

This gives us an uncertainty of 2.46 km s$^{-1}$ on either side of our velocity estimate with corresponding uncertainty in kinematic distance of 0.2 kpc.

### 4.2. Linear Polarization and Intrinsic RM

As mentioned in Section 3.2, we note a curious feature in the polarization intensity maps, in which there is an abrupt break in the southwest part of the shell where a depolarized "strip" cuts across.

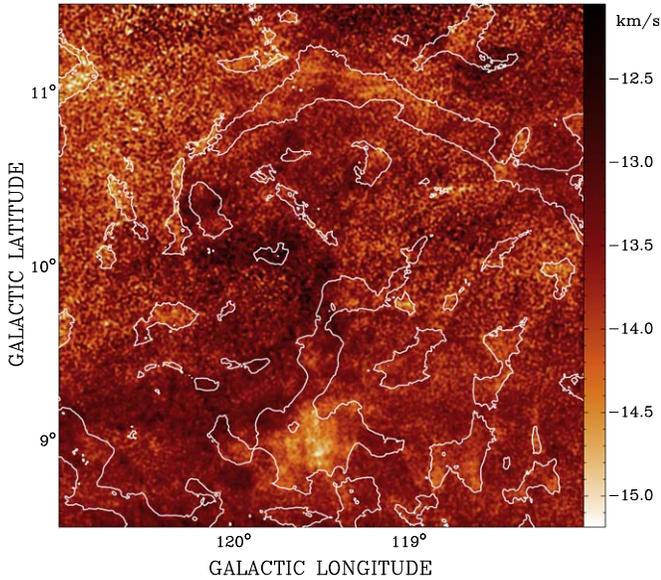

**Figure 6.** First moment map of the H I emission made with the velocity range containing the H I shell continuation (see Figure 4). The color intensity corresponds to H I velocity in km s$^{-1}$ and the contours overlaid in white show the outline of the RHT-identified filamentary structure for that velocity range.

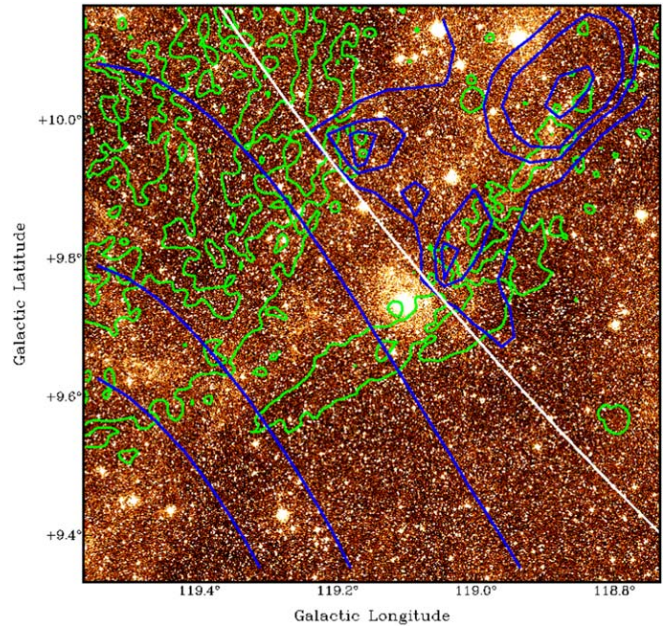

**Figure 7.** Optical emission at wavelength 0.67 μm from the Digitized Sky Survey with linear polarization contours overlaid in green. Blue contours indicate H$_\alpha$ emission taken from D. P. Finkbeiner (2003). The white line marks the edge of the Virginia Tech Sky Survey (VTSS) coverage.

We see no evidence of this strip in the total power continuum images, so this is likely caused by something in the foreground, although we can not conclusively determine a cause. In Figure 7, we show a red optical Digitized Sky Survey (DSS-red) image of the area. We see that the gap in the southern part of the SNR shell aligns well with faint optical background emission, which splits off at the bottom of the shell with part of it continuing to the south and part of it continuing along the curvature of the shell toward the east. This could be indicative of filaments of ionized gas in the foreground of the SNR, visible only at optical wavelengths as we did not see evidence of these filaments in the radio continuum or H I emission. If this indeed is ionized gas, the optical emission we detect should be H$_\alpha$ recombination line emission, which falls into the red frequency band displayed in Figure 7. Unfortunately, this area of space has not been covered by the H$_\alpha$ Virginia Tech Sky Survey as can be seen in Figure 7. However, the continuation of this red filament in the DSS-red to the top right coincides with H$_\alpha$ emission seen in the D. P. Finkbeiner (2003) map displayed in Figure 7 as blue contours. This is an indication that the DSS-red filament coinciding with the gap in polarization consists of H$_\alpha$ emission and therefore contains ionized gas. We did not see a coincidence of the optical line emission published by F. Mavromatakis et al. (2000) with the polarization gap.

Parts of the optical filament seem to follow the same curvature as the radio shell and do not depolarize the radio emission. It is therefore plausible to assume that the source of the optical emission is related to the SNR itself or the wind of the progenitor star. Since this is a three-dimensional object, there could be a very complex geometry with the depolarizing filament being in front of the radio shell in some parts and behind it in other parts of the SNR. This may indicate that the depolarizing filaments are part of the SNR complex.

We also attempt to address the claim made in X. H. Sun et al. (2011) that there is a lens or Faraday screen in the foreground with RM of $\approx +45$ rad m$^{-2}$, partially blocking CTA 1. X. H. Sun et al. (2011) present a map of extragalactic sources over a large field (115°–125° Galactic longitude, 7°–13° Galactic latitude) around CTA 1 to demonstrate that their proposed Faraday screen extends beyond the SNR, with a radius of about 3°. We reproduce this map in Figure 8, using an extensive RM catalog (C. L. Van Eck et al. 2021) that includes the catalog derived from the NRAO Very Large Array Sky Survey (A. R. Taylor et al. 2009) used by X. H. Sun et al. (2011), but also includes newer catalogs, most notably the Canadian Galactic Plane Survey RM catalog (C. L. Van Eck et al. 2021), which partly overlaps with CTA 1. In Table 2 we list all 30 compact polarized sources in the area we observed around CTA 1 for which we were able to determine RMs. All of the previously published sources are detected in our observations, but three have a percentage polarization below 2%, which makes the calculation of RMs unreliable. A total of 14 of the 30 sources are new detections in our data. We see the same cluster of sources with positive RMs near the side of the shell with RMs of the same sign, but we find additional sources in our new measurements with negative RMs that are intermixed with the compact sources showing the positive RMs. Therefore, the foreground lens proposed by X. H. Sun et al. (2011) is not as clearly outlined as it was before. We also conducted a preliminary search for the proposed Faraday screen in the data from the Global Magneto-Ionic Medium Survey (GMIMS; M. Wolleben et al. 2019). The observations from GMIMS span 1000 frequency channels and the resolution is 35′, so if there were a Faraday screen in the foreground of CTA 1, we would expect to see evidence of the screen affecting emission in that region. However, we saw no such evidence.

We also attempted to follow X. H. Sun et al.'s (2011) method of estimating intrinsic RM from the relative depolarization between observed wavelengths, first by combining the ST's continuum bands to compare wavelengths $\lambda_{AB} = 21.26$ cm and $\lambda_{CD} = 20.96$ cm, and then by comparing just the highest and lowest bands with wavelengths $\lambda_A = 21.30$ cm and $\lambda_D = 20.90$ cm. However, the signal-to-noise ratios for our depolarization maps were not high enough to be confident in our analysis, so we have not included our depolarization or intrinsic RM maps.

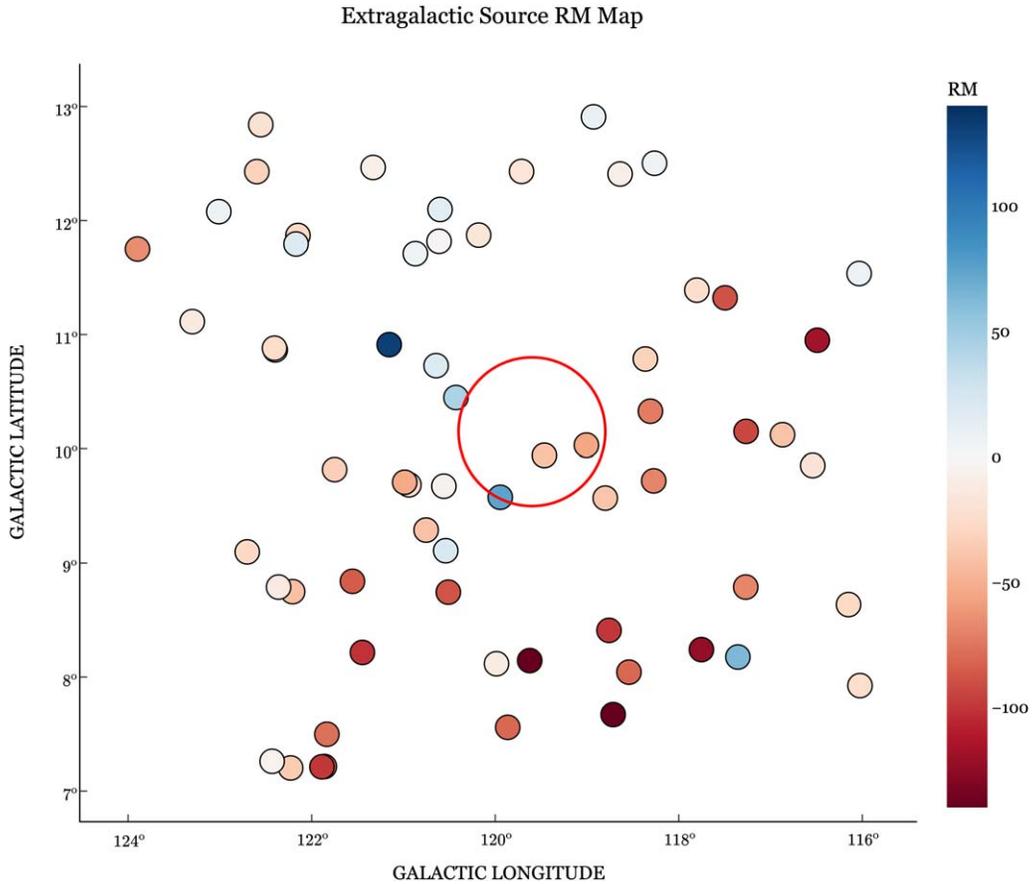

**Figure 8.** RM map of extragalactic sources from the catalogs compiled by C. L. Van Eck et al. (2021) and A. R. Taylor et al. (2009) in the area surrounding CTA 1. We also added compact source RMs determined with our observations (see Table 2). Each marker is colored by RM with red coloring indicating negative values, while positive values are colored in blue. The large red circle indicates the approximate region that CTA 1 spans. The values of the RM magnitudes in this figure range from −186 to +132 rad m$^{-2}$.

With more planned ST observations at higher Galactic latitudes (for better coverage of the region where the proposed Faraday screen is suggested to be), we wait to integrate the ST data with that of the GMIMS data and more confidently address the possibility of a Faraday screen in a future paper.

However, we do suggest another possible interpretation of the difference in sign for RMs across the shell, inspired by a similar investigation into Faraday rotation of an SNR, in which L. Harvey-Smith et al. (2010) propose that the source of this difference is a swept up stellar wind from the progenitor star, with a toroidal magnetic field. In this interpretation, we would expect to see a consistent magnitude of RM across the shell, with opposite sign across a line of symmetry.

Given this expectation, we can make a rough estimate for the Galactic foreground contribution to RM by determining the average observed RM for the shell on either side of the line of symmetry. Since the observed RM is the addition of the Galactic foreground RM and the intrinsic RM of the SNR, subtracting the same RM from the average observed RM on both sides of the shell should result in RMs of the same magnitude but opposite sign. We were unable to create an observed RM map with our diffuse polarization data from the ST because our single antenna data include only one frequency (1420 MHz) and as mentioned, we wait for more fields toward the northeast part of the shell to be observed before combining the GMIMS data to create a high-resolution diffuse RM map. Instead, we used the RM values based on the RM map displayed in X. H. Sun et al. (2011).

CTA 1 shows a bilateral symmetry, as known from many SNRs. The angle of the symmetry axis with the Galactic plane is about 54° (J. L. West et al. 2016). Following the line of thought of L. Harvey-Smith et al. (2010), we plotted RM as a function of distance from that symmetry axis in Figure 9. We find an average of about +7 rad m$^{-2}$ on the eastern shell and −42 rad m$^{-2}$ on the western shell. Correcting each shell for a Galactic foreground contribution of −17.5 rad m$^{-2}$ results in an average absolute intrinsic RM of about 24.5 rad m$^{-2}$. This is consistent with the estimate of the intrinsic RM using the method of relative depolarization between bands and the Galactic foreground RM value of −17 ± 8 rad m$^{-2}$ calculated by X. H. Sun et al. (2011) using the three-dimensional emission models of the Milky Way by X. H. Sun et al. (2008) and X.-H. Sun & W. Reich (2010).

Using their Equation (10) and adding the angle between the line of sight and the plane of the toroidal magnetic field Θ back in, we get for diffuse emission,

$$RM = 0.9 \left(\frac{R}{\Delta R}\right)^{\frac{3}{2}} \cos \Theta \text{ rad m}^{-2}, \quad (2)$$

accounting for the factor of 2 between the measured RM inside the SNR shell and the RM that would be seen by a point source

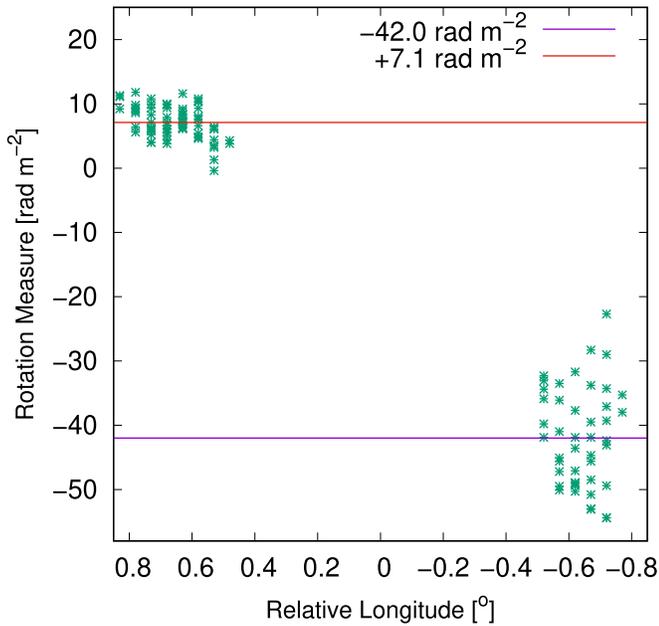

**Figure 9.** Plots of RM as a function of angular distance from the symmetry axis (see Figure 1). Only RMs outside of 0°.5 are shown to avoid the polarized emission from the interior of the SNR. The average RM values of $+7.1$ rad m$^2$ for the eastern shell and $-42$ rad m$^2$ for the western shell are indicated by the red and purple lines, respectively.

**Table 2**
List of Compact Source RMs in Our Observations

| Galactic Longitude (deg) | Galactic Latitude (deg) | Catalog RMs | New RMs | %-Pol. (%) |
|---|---|---|---|---|
| 117.266 | 10.151 | ⋯ | $-93 \pm 27$ | 3.9 |
| 117.270 | 8.789 | $-68.7$ | ⋯ | 1.6 |
| 117.355 | 8.176 | ⋯ | $+63 \pm 25$ | 3.0 |
| 117.495 | 11.322 | ⋯ | $-89 \pm 15$ | 10.5 |
| 117.753 | 8.241 | ⋯ | $-124 \pm 26$ | 28.2 |
| 117.803 | 11.390 | $+17.3$ | $-24 \pm 4$ | 5.8 |
| 118.273 | 9°.721 | ⋯ | $-69 \pm 53$ | 17.6 |
| 118.309 | 10.328 | $-16.1$ | $-72 \pm 18$ | 5.5 |
| 118.364 | 10.786 | $-22.7$ | $-32 \pm 23$ | 3.6 |
| 118.543 | 8.043 | $-98.7$ | $-81 \pm 36$ | 3.6 |
| 118.715 | 7.671 | ⋯ | $-186 \pm 22$ | 3.5 |
| 118.760 | 8.410 | ⋯ | $-100 \pm 14$ | 7.5 |
| 118.801 | 9.570 | $-9.8$ | $-39 \pm 8$ | 7.1 |
| 119.006 | 10.031 | $-27.3$ | $-55 \pm 19$ | 7.6 |
| 119.464 | 9.942 | $-32.9$ | $-41 \pm 12$ | 12.2 |
| 119.624 | 8.144 | ⋯ | $-175 \pm 45$ | 2.9 |
| 119.945 | 9.577 | ⋯ | $+75 \pm 17$ | 10.5 |
| 119.986 | 8.117 | ⋯ | $-12 \pm 22$ | 2.7 |
| 120.428 | 10.448 | $+2.5$ | $+44 \pm 43$ | 3.6 |
| 120.512 | 8.745 | $-42.3$ | $-89 \pm 66$ | 3.1 |
| 120.540 | 9.108 | $+22.2$ | ⋯ | 0.7 |
| 120.560 | 9.674 | $+19.7$ | $-7 \pm 5$ | 2.9 |
| 120.644 | 10.727 | $+19.3$ | ⋯ | 0.3 |
| 120.753 | 9.290 | ⋯ | $-41 \pm 9$ | 8.2 |
| 120.940 | 9.687 | $+2.6$ | $-26 \pm 14$ | 6.5 |
| 120.987 | 9.709 | $-1.5$ | $-54 \pm 45$ | 7.1 |
| 121.155 | 10.911 | $+3.6$ | $+132 \pm 55$ | 4.9 |
| 121.447 | 8.216 | ⋯ | $-102 \pm 59$ | 6.2 |
| 121.556 | 8.840 | ⋯ | $-85 \pm 9$ | 5.9 |
| 121.750 | 9.820 | ⋯ | $-34 \pm 26$ | 11.9 |

**Note.** RMs are given in rad m$^{-2}$.

behind it (B. J. Burn 1966). In our case, with $|RM| = 24.5$ rad m$^{-2}$, we require a ratio between the SNR shell's radius $R$ and its thickness $\Delta R$ of 9.0. CTA 1 is supposed to be in the adiabatic or Sedov–Taylor expansion phase, for which this ratio is expected to be 10 (L. I. Sedov 1959). Using this ratio in the equation would lead to $\Theta \approx 30°$, a perfectly reasonable result.

We conclude that the results of our radio continuum and polarization observations of CTA 1 are consistent with an SNR expanding inside the wind zone of a red supergiant.

## 5. Conclusion

We have presented the most recent DRAO ST radio observations of CTA 1 in the 1420 and 408 MHz continuum and linear polarization at 1420 MHz, as well as the H I environment around the SNR. In addition to confirming the previously determined continuum features from S. Pineault et al. (1997), we describe the filamentary structure of the shell that is visible in our newer continuum map at 1420 MHz. In analyzing the linearly polarized emission from CTA 1 at 1420 MHz, we note a depolarized strip through the shell, which is of particular interest, as there is no indication of such a gap in either of the continuum maps. Investigating this gap beyond the radio continuum, we find that an optical emission map from the Digitized Sky Survey seems to show a filament that is aligned with the depolarized strip, indicating possible depolarizing gas visible only at optical wavelengths. As parts of these optical filaments also share a similar shape and overlap with part of the SNR shell, we suggest that the source of the optical emission could be leftover ionized gas from the progenitor star, produced before the explosion. We also analyze the linear polarization data using the method of estimating intrinsic RM from relative depolarization described in X. H. Sun et al. (2011), and find high relative depolarization corresponding to intrinsic RM values mostly in the $10 - 31$ rad m$^{-2}$ range with higher values ($\approx 60$ rad m$^{-2}$) in the central emission bridge/western shell regions, consistent with X. H. Sun et al.'s (2011) findings. Following their line of investigation, we created a similar RM map of extragalactic sources around CTA 1 with an extended catalog from C. L. Van Eck et al. (2021) and confirm that the sign of the RMs for extragalactic sources near and around the SNR shell match that of the shell itself. However, a cursory look through GMIMS data T. L. Landecker et al. (2010) to search for X. H. Sun et al.'s (2011) proposed foreground Faraday screen prompted us to consider other alternatives to the cause of the abrupt RM sign change across the shell and extragalactic sources, as we could not see any obvious evidence of a foreground lens. With future ST observations planned and further analysis of the GMIMS T. L. Landecker et al. (2010) data, we expect a more conclusive confirmation or rejection of the possibility of this foreground lens in a future paper. A promising possible alternative to the existence of a screen is that the difference in sign could be due to the SNR's expansion in a swept-up stellar-wind zone from the progenitor star. L. Harvey-Smith et al. (2010) showed for another SNR, G296.5 +10.0, that a red supergiant star, with its rotation axis along the symmetry axis of the SNR shell, could cause a stellar wind with a toroidal magnetic field that produces the RM sign flip across the SNR shell symmetry axis. Finally, we demonstrate the use of RHT analysis on H I around an SNR to identify the filamentary structure and better identify associated H I features to make more accurate kinematic distance estimates. Using an updated systemic velocity of $-13.8 \pm 2.46$ km s$^{-1}$ for the shell, we estimate a distance of $1.22 \pm 0.2$ kpc using the IAU standard Galactic rotation curve and a distance of $1.09 \pm 0.2$ kpc using the Galactic rotation curve from M. J. Reid et al. (2019).

### Acknowledgments

We would like to thank Rohit Dokara for the careful reading of the manuscript. A.S.H. is supported by an NSERC Discovery Grant. The Dominion Radio Astrophysical Observatory is a National Facility operated by the National Research Council Canada. The Digitized Sky Surveys were produced at the Space Telescope Science Institute under U.S. Government grant NAG W-2166. The images of these surveys are based on photographic data obtained using the Oschin Schmidt Telescope on Palomar Mountain and the UK Schmidt Telescope. The plates were processed into the present compressed digital form with the permission of these institutions. The Virginia Tech Spectral-Line Survey is supported by the National Science Foundation.

*Facility:* DRAO:Synthesis Telescope.

### ORCID iDs

Roland Kothes  https://orcid.org/0000-0001-5953-0100
Alex S. Hill  https://orcid.org/0000-0001-7301-5666
Andrew Gray  https://orcid.org/0000-0002-2280-7644
Wolfgang Reich  https://orcid.org/0000-0002-5313-6409